\documentclass[amsmath,amssymb,pra,aps,showpacs,twocolumn,10pt,superscriptaddress]{revtex4-1}
\usepackage{amsmath,amsfonts,amssymb,amsthm,graphics,graphicx,epsfig,bbm}
\usepackage[colorlinks=true,citecolor=blue,linkcolor=red,urlcolor=blue]{hyperref}
\usepackage[usenames]{color}
\usepackage{graphicx}
\usepackage{subfigure}
\usepackage{amsmath}
\usepackage{epsfig}
\usepackage{dcolumn}
\usepackage{bm}
\usepackage{color}
\usepackage{epstopdf}
\usepackage{amssymb}
\usepackage{amstext}
\usepackage{latexsym}
\usepackage{hyperref}
\usepackage{amsfonts}
\usepackage{psfrag}
\usepackage{xcolor}
\usepackage[normalem]{ulem}
\usepackage{dsfont}
\usepackage{txfonts}
\usepackage{overpic}
\usepackage{ifthen}
\usepackage{comment}

\newcommand{\Tr}{\mathrm{Tr}}

\newcommand{\an}[2]{\ifthenelse{\equal{#1}{}}{\ensuremath{\hat{#1}_{#2}}}{\ensuremath{\hat{#1}^{\protect\phantom{\dagger}}_{#2}}}}

\begin{document}
\title{Dephasing-induced growth of discrete crystalline order in spin networks}

\author{Akitada Sakurai}
\email{akitada-phy@nii.ac.jp}
\affiliation{Department of Informatics, School of Multidisciplinary Sciences, SOKENDAI (The Graduate University for Advanced Studies), 2-1-2 Hitotsubashi, Chiyoda-ku, Tokyo 101-8430, Japan}
\affiliation{National Institute of Informatics, 2-1-2 Hitotsubashi, Chiyoda-ku, Tokyo 101-8430, Japan}

\author{Victor M.~Bastidas}
\affiliation{NTT Basic Research Laboratories \& Research Center for Theoretical Quantum Physics,  3-1 Morinosato-Wakamiya, Atsugi, Kanagawa, 243-0198, Japan} 
\affiliation{National Institute of Informatics, 2-1-2 Hitotsubashi, Chiyoda-ku, Tokyo 101-8430, Japan}

 \author{Marta P.~Estarellas}
 \altaffiliation[Contact Address: ]{Qilimanjaro Quantum Tech., Carrer dels Comtes de Bell-Lloc, 161, 08014 Barcelona, Spain}
 \affiliation{National Institute of Informatics, 2-1-2 Hitotsubashi, Chiyoda-ku, Tokyo 101-8430, Japan}
 
\author{William J.~Munro}
\affiliation{NTT Basic Research Laboratories \& Research Center for Theoretical Quantum Physics,  3-1 Morinosato-Wakamiya, Atsugi, Kanagawa, 243-0198, Japan} 
\affiliation{National Institute of Informatics, 2-1-2 Hitotsubashi, Chiyoda-ku, Tokyo 101-8430, Japan}

\author{Kae Nemoto}
\affiliation{National Institute of Informatics, 2-1-2 Hitotsubashi, Chiyoda-ku, Tokyo 101-8430, Japan}
\affiliation{Department of Informatics, School of Multidisciplinary Sciences, SOKENDAI (The Graduate University for Advanced Studies), 2-1-2 Hitotsubashi, Chiyoda-ku, Tokyo 101-8430, Japan}

%%%%%%%%
\begin{abstract}
%%%%%%%%
A quantum phase of matter can be understood from the symmetry of the system's Hamiltonian. The system symmetry along the time axis has been proposed to show a new phase of matter referred as discrete-time crystals (DTCs).  A DTC is a quantum phase of matter in non-equilibrium systems, and it is also intimately related to the symmetry of the initial state. DTCs that are stable in isolated systems are not necessarily resilient to the influence from the external reservoir. 
In this paper, we discuss the dynamics of the DTCs under the influence of an environment. Specifically, we consider a non-trivial situation in which the initial state is prepared to partly preserve the symmetry of the Liouvillian.  Our analysis shows that the entire system evolves towards a DTC phase and is stabilised by the effect of dephasing.  Our results provide a new understanding of quantum phases emerging from the competition between the coherent and incoherent dynamics in dissipative non-equilibrium quantum systems.
%%%%%%%%
\end{abstract}
%%%%%%%%

\maketitle
\date{\today}

\section{Introduction}

Understanding phases of matter and realizing transitions between them have been a central theme in quantum many-body physics~\cite{Yang54, Cooper56, Higgs1964}.   Quantum phases of matter are a macroscopic property, which reflects the underlying microscopic structure of the system~\cite{landau1981}.  In closed quantum systems, for example, this microscopic structure is described in terms of the Hamiltonian, whose properties and symmetries determine the quantum phases that the system could exhibit~\cite{Cooper56}.  
As the phases of matter are intimately related to symmetries of the system, they can be quite rich, ranging from ferromagnetism to superconductivity~\cite{landau1981}.

A discrete-time crystal (DTC) is a quantum phase of matter related to the symmetries of a periodically-driven system; it appears when the discrete time-translational symmetry (DTTS) of the Hamiltonian is broken~\cite{Else2016, Else2017, Sacha2015, Sacha2018, Khemani2016, Berdanier2018, Giergiel2018, Guo2020, Guo2013, Guo2016, Russomanno2017, Pizzi2019, Pizzi2020, Pizzi2021, Fan2020, Surace2019, Iemini2018, Gambetta2019, Andreu20, Estarellas19, Bastidas2020, Sakurai2021, Yu2019, Kozin2019, Choi2017, Zhang2017, Ho2017}.  In non-equilibrium systems, the symmetry breaking can be caused by parameter changes in the Hamiltonian or by preparing particular initial states that break the symmetries of the system~\cite{Else2016, Else2017}.  
The latter can be interpreted as a quantum quench~\cite{Polkovnikov2011}, because to generate the DTC phase the system has to be prepared in an appropriate symmetry-broken state and not in one of its eigenstates.   

As quantum phases and the system symmetries are closely related, it is known that when a system is coupled to an external environment, some of its symmetries can be broken and the phases of matter associated to them do not survive. One example of this is a Mott insulator at unit filling coupled to a zero temperature bath~\cite{Boite2014}. In this situation, the coupling to the bath breaks the conservation of particles and the Mott insulator phase is destroyed.
When a system is coupled to a Markovian environment, quantum phases of matter can be well defined in terms of symmetries of the Liouvillian~\cite{Albert2014,Tangpanitanon2019}.
In the case of DTC's, the coupling to an external environment can destroy the discrete time-crystalline order~\cite{Lazarides17, Fan2020}.  To prevent this situation, the Liouvillian has to satisfy the relevant symmetries associated to the DTC phase in the closed system~\cite{Andreu20, Iazarides2020}. Further, the steady state should break the aforementioned symmetries giving rise to the DTC. Recently, a DTC phase in a dissipative atom-cavity system has been theoretically predicted~\cite{kessler2020continuous} and has been experimentally realized using a Bose Einstein condensate within a cavity~\cite{Hans20}.

In this paper, we explore the competition between the local symmetries of the initial state and the global symmetries of the Liouvillian operator, and we investigate the dynamics of the system before it reaches the steady state. Global symmetries of the Liouvillian allow us to define conserved quantities, which in turn determine properties of the steady state~\cite{Albert2014}. However, not all the symmetries of the closed system are available when we coupled the system to an environment~\cite{Tangpanitanon2019}. If we prepare an initial state breaking a symmetry of the Hamiltonian, as well as the symmetry of the Liouvillian, the quantum phase could survive at the steady state under certain circumstances~\cite{Minganti2018}. Here we explore a situation where two different phases of matter with different symmetries are initially prepared in well-defined regions of space. One region of the system is in a state that locally breaks a symmetry of the Liouvillian, while the other region preserves it. In this situation, it is not clear what happens to these phases during the time evolution before the system reaches its steady state.

\begin{figure}[t]
\centering
\includegraphics[width=0.45\textwidth]{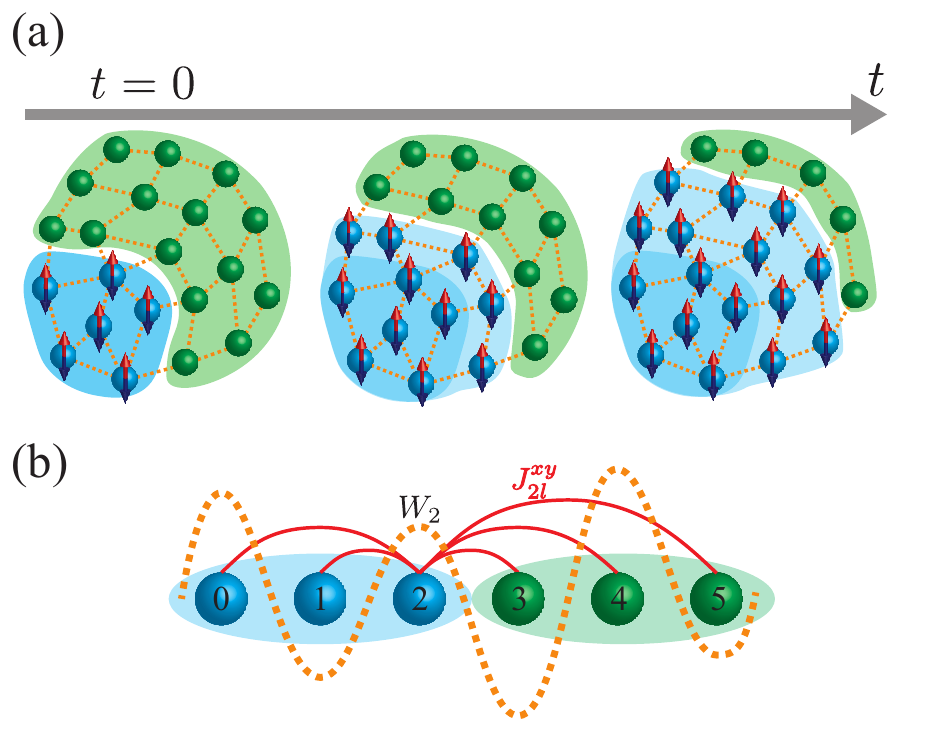}
\caption{ (a) Schematic illustration showing the concept of the transformation of two quantum phases in time. The left panel shows the initial state with two distinctive quantum phases in a spin network.  In the region A (blue), the initial state breaks breaks the discrete time translational symmetry and the initial state in the region B (green) does not break the symmetry.  (b) Diagram of our system for a spin network of $N=6$ sites. Here, the red lines represent the the long range interaction $J_{lm}^{xy}$ between sites, and the dashed line means diagonal disorder $W_l$ at the site. The geometrical arrangement of spins is one-dimensional, but they have long range interactions determining the connectivity of the spin network.  We label spins to form the region A with the sites $l=0,1,2$, the region B with the sites $l=3,4,5$. 
}
\label{Fig1}
\end{figure}

To investigate this more closely, we independently prepare an initial state to give a DTC phase or a non-DTC phase for each well-defined region of space, as shown in Fig.~\ref{Fig1} (a) with the DTC region (blue) and non-DTC region (green). These two phases are accommodated in a spin network shown in Fig. \ref{Fig1} (b).  Then, we evaluate the transformation of these quantum phases in time through a dissipative process. In the next section, we start with defining our system.

\section{Model}
In our work we will focus on a DTC, a phase of matter that breaks the discrete time translational invariance of the Hamiltonian. To define this phase, we consider a periodically driven $N$-site spin network governed by the Hamiltonian
\begin{equation}
    \hat{H}(t) = 
    \begin{cases}
    \hat{H}_1 =\hbar g(1-\epsilon) \sum\limits_{l}  \sigma_{l}^{x}  & 0 \leq t < T_1 \\
    \hat{H}_2 =\hbar \sum\limits_{lm} J_{lm}^{xy} \left( \sigma_{l}^{x} \sigma_{m}^{x} + \sigma_{l}^{y} \sigma_{m}^{y}  \right) + \hbar \sum\limits_{l}W_l^z \sigma_l^z & T_1 \leq t < T \ ,
    \end{cases}
    \label{eq:Hamiltonian}
\end{equation}
where $T=T_1+T_2$ is the period and $\sigma^{\mu}$ ($\mu \in x,y,z$) are the Pauli operators acting on the $l$-th site. Further we set $T_1 =T_2 = T/2$ with $2gT_1= \pi$. Next, $\epsilon$ is a rotational error parameter, which indicates the deviation of the Hamiltonian from the perfect DTC case. $W_l$ denotes the on-site energy at the $l$-th site and is given by a uniform random distribution $W_l^z  \in [0,W]$ with disorder strength $W$.  Unless explicitly stated, $\epsilon$ and $W$ are set to zero in this paper.  This $N$-site spin system can be considered as a network, the sites being its nodes and the coupling $J_{lm}^{xy}$ between the sites $l$ and $m$ being the edges of the network. Crucially, in the absence of error ($\epsilon=0$), the evolution operator $\hat{U}(t)$ associated to Eq.~\eqref{eq:Hamiltonian} preserves the total number of excitations $\hat{\mathcal{N}}=\sum_l \left(\sigma_l^z+1\right)/2$ at even periods that is $[\hat{U}(2nT),\hat{\mathcal{N}}]=0$. The consequence of this symmetry is that the dynamics will preserve the number of excitations of the initial state at stroboscopic times. The physical origin of this stroboscopic conserved quantity is clearer if we examine the effective Hamiltonian $\hat{H}^{\text{eff}}_{2T}$ after two periods of the drive in the absence of disorder $WT/2\pi=0$ and for $\epsilon=0$. The effective Hamiltonian is defined in terms of the unitary operator after two periods $\hat{U}(2T)=\exp{\left(-2i\hat{H}^{\text{eff}}_{2T}T\right)}$, giving
\begin{equation}
   \hat{H}^{\text{eff}}_{2T,W=0}=
    \frac{\hbar}{2} \sum\limits_{lm} J_{lm}^{xy} \left( \sigma_{l}^{x} \sigma_{m}^{x} + \sigma_{l}^{y} \sigma_{m}^{y}  \right)  
   \ .
    \label{eq:2TeffHamiltonian}
\end{equation}
Here the factor $1/2$ in the second term comes from a factor $T_2/T$. Here, the above Hamiltonian is invariant under the global $U(1)$ symmetry, which leads to the preservation of the total number of excitations. In particular, this Hamiltonian has a $\mathbb{Z}_2$-Ising symmetry (the invariant under the transformation $\sigma^y_l\mapsto -\sigma^y_l$ and $\sigma^z_l\mapsto -\sigma^z_l$), and thus the order parameter of the our DTC is the local magnetization $\langle \sigma_l^z (nT) \rangle $ at periodic times ($t=nT$)~\cite{Else2017}.   When the initial state breaks the $\mathbb{Z}_2$-Ising symmetry, the magnetization oscillates with a $2T$-periodicity, while preserving the total number of excitations.

\subsection{Lindblad equation}
In this paper, we assume that our spin network is weakly coupled to a Markovian environment~\cite{Gardiner2004}.
In many of solid-state quantum systems, dephasing is the dominant dissipative factor and we consider the Hamiltonian (\ref{eq:Hamiltonian}) with dephasing.  The dynamics of the system is given by the Lindblad equation~\cite{Gorini1976, Lindblad1976, Scopa2018, Manzano2020}
\begin{equation}
    \frac{\partial}{\partial t} \hat{\rho}(t) = \hat{\mathcal{L}}_t[\hat{\rho}(t)]
    \ ,
    \label{eq:Lindblad}
\end{equation}
where $\hat{\rho}$ is  the density matrix of the spin network and $\hat{\cal{L}}_t$ the Liouvillian operator.  The action of $\hat{\mathcal{L}}_t$ on the reduced density matrix of the system $\hat{\rho}(t)$ is given by
\begin{equation}
	 \hat{\mathcal{L}}_t[\hat{\rho}(t)] = - \frac{i}{\hbar} \left[ \hat{H}(t),\hat{\rho}(t) \right] + {\gamma} \sum_l \left[\sigma_l^z \hat{\rho}(t)\sigma_l^z - \hat{\rho}(t)\right],
\label{eq:Linovillian}
\end{equation}
where $\gamma$ is the dephasing rate. As the Hamiltonian is periodic in time $\hat{H}(t)=\hat{H}(t+T)$, the Liouvillian operator exhibits the same periodicity such that $\hat{\mathcal{L}}_t=\hat{\mathcal{L}}_{t+T}$~\cite{Dai2016, Scopa2018}. In our model, we assume that during the spin rotation given by $\hat{H}_1 $, the dephasing effect can be neglected~\cite{Else2017, Andreu20}, because in many experiments the pulse can be applied in short time which is much smaller than the time scale of the dephasing~\cite{Choi2017, Zhang2017, Hans20, Lanyon57, Cirac1995, Blatt2012}.

The Liouvillian in Eq. \eqref{eq:Lindblad} is the generator of a completely positive map that preserves the trace of the density matrix, and it is also a map between the vector space of the linear operators. To analyze the properties of this map,  we apply the super-operator formalism~\cite{Manzano2020, Pablo2004}. The density matrix $\hat{\rho}$ is then written as a vector $ |\hat{\rho} \rangle\rangle = \rho_{lm} |l \rangle \otimes |m \rangle$  in an extended Hilbert space $\mathcal{H} \otimes \mathcal{H}$, where $|l\rangle$ and $|m\rangle$ are the basis states of the Hilbert space $\mathcal{H}$ and $\rho_{lm} = \langle l |\hat{\rho}| m \rangle$. In this extended Hilbert space, the inner product of two operators $\hat{A}$ and $\hat{B}$ is defined as $\langle\langle \hat{A}|\hat{B} \rangle\rangle = \Tr{(\hat{A}^\dagger \hat{B})}$. 

Now, the Lindblad equation \eqref{eq:Lindblad} can be written as a linear system of coupled ordinary differential equations~\cite{Manzano2020}
\begin{equation}
	\frac{\partial}{\partial t} |\hat{\rho}(t) \rangle\rangle = \hat{\hat{L}}_t |\hat{\rho}(t) \rangle\rangle \ .
		\label{eq:SuperLindblad}
\end{equation}
In this formalism, the Liouvillian super-operator $\hat{\hat{L}}_t$ is given by
\begin{equation}
   \hat{\hat{L}}_t =  - \frac{i}{\hbar} \left(\hat{H}\otimes \mathbb{I}_N - \mathbb{I}_N \otimes \hat{H}\right) +  \gamma \sum_l \left(  \sigma_l^z\otimes\sigma_l^z - \mathbb{I}_N\otimes\mathbb{I}_N \right),
\end{equation}
with $\mathbb{I}_N$ being the $2^N \times 2^N$-dimensional identity operator. 

\begin{figure*}[t]
\centering
\includegraphics[width=0.96\textwidth]{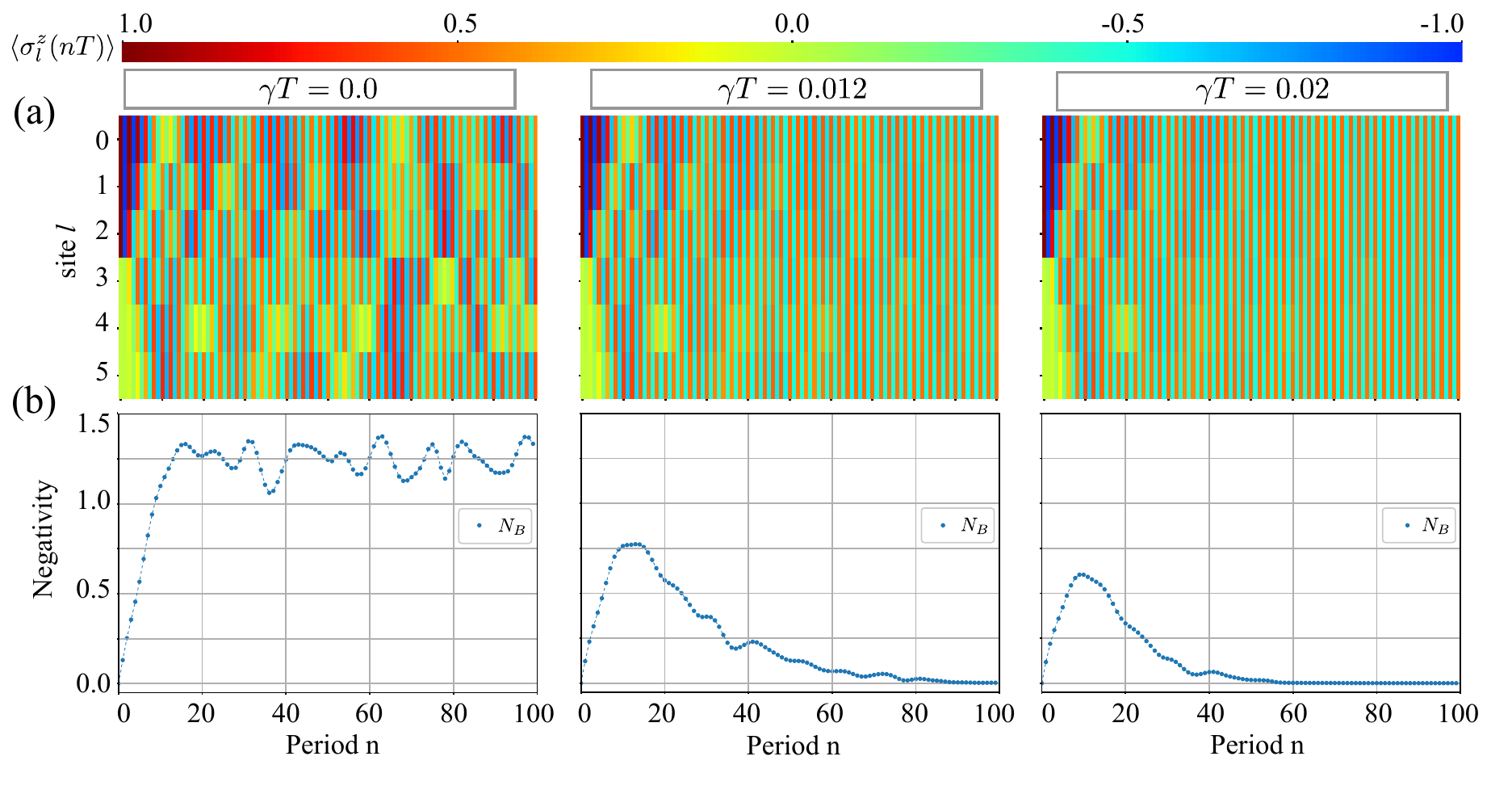}
\caption{  Stroboscopic dynamics of the local magnetization $\langle \sigma_l^z(nT) \rangle$ and the negativity $N_B (nT)$ for DTC growth in the different dephasing rates $\gamma T =0, 0.012, 0.02$ are shown in (a) and (b), respectively. Here, we have chosen, $J_0 T/2\pi =0.2$ with $\alpha=1.5$ and $  WT/2\pi=0.0$. We have used a state $|\psi(0)\rangle = |111\rangle_A \otimes |+++\rangle_B$ as our initial state.
}
\label{Fig2}
\end{figure*}

\subsection{Floquet theory}
As a DTC is a phase of matter that can be defined in periodically-driven systems, the Floquet theory is a convenient and versatile tool to represent and analyze its dynamics.  Floquet theory is related to the study of linear systems of coupled differential equations with time-periodic coefficients and it can be naturally used to investigate the dynamics governed by Eq.~\eqref{eq:SuperLindblad} because $\hat{\hat{L}}_t=\hat{\hat{L}}_{t+T}$.  The evolution at stroboscopic times $t_n=nT$ is given by $|\hat{\rho} (nT) \rangle \rangle = \hat{\hat{\Phi}}_{T}^n |\hat{\rho}(0)\rangle\rangle$ with the dynamical map $\hat{\hat{\Phi}}_{T}$~\cite{Dai2016, Scopa2018}
\begin{equation}
	\hat{\hat{\Phi}}_{T} =\hat{\mathcal{T}}_{\leftarrow}\,e^{\int_0^T \hat{\hat{L}}_{\tau} d\tau} = e^{\hat{\hat{L}}^{\text{eff}}_{T}T },
	\label{eq:effLiouvillian}
\end{equation}

where $\hat{\mathcal{T}}_{\leftarrow}$ is the time-ordering operator (right to left). The dynamics at stroboscopic times $t_n=nT$ are governed by the effective Liouvillian super-operator $\hat{\hat{L}}^{\text{eff}}_{T}  = \log{\hat{\hat{\Phi}}_T}/T$.  In the absence of error $(\epsilon=0)$, the dynamical map preserves some symmetries of the closed system. For example, it preserves the total number of excitations every two periods in such a way that $[\hat{\hat{\Phi}}^{2n}_{T},\hat{\hat{\mathcal{N}}}]=0$, where $\hat{\hat{\mathcal{N}}}$  is the super operator associated to $\hat{\mathcal{N}}$. This symmetry allows us to decompose even powers of the dynamical map in blocks with a different number of excitations. The choice of the initial state determines the total number of excitations of the system and how many symmetry multiplets are relevant for the dynamics. This in turn determines the properties of the steady state. It is instructive to investigate the origin of the aforementioned conserved quantity. With this aim, let us consider the system in the absence of disorder and error. In this case, we can explicitly obtain the effective Liouvillian operator associated to $\hat{\hat{\Phi}}^{2}_{T}$, as follows
\begin{equation}
   \hat{\hat{L}}^{\text{eff}}_{2T} =   - \frac{i}{\hbar}  \left( \hat{H}^{\text{eff}}_{2T}\otimes \mathbb{I}_N - \mathbb{I}_N \otimes \hat{H}^{\text{eff}}_{2T} \right) +  \frac{\gamma}{2} \sum_l \left(  \sigma_l^z\otimes\sigma_l^z - \mathbb{I}_N\otimes\mathbb{I}_N \right).
   \label{eq:a-2TeffLiouvillian}
\end{equation}
Here the factor $1/2$ in the second term comes from a factor $T_2/T$. Due to the nature of the dephasing and the conservation rules of the effective Hamiltonian $\hat{H}^{\text{eff}}_{2T}$, it is clear that the dissipative process also preserves the total number of excitations $\hat{\mathcal{N}}$.

\section{Results}
We use the model defined in the previous section to numerically analyze dynamics of the phases initially prepared in two domains of the system and to investigate its dynamics.  Firstly, we fix the spin network properties by setting the coupling strength.  We set each coupling strength between two spins as $J_{lm}^{xy} = J_0/|l-m|^\alpha$ where the index $l$ and $m$ denotes the sites.  $J_0$ and $\alpha$ denote the strength and range of the spin-spin interaction, respectively.  
In the limiting cases $\alpha=0$ and $\alpha=\infty$, the connectivity of the network is all-to-all  and the nearest neighbor only respectively. In this paper, we choose $\alpha = 1.51$, which is an experimentally used value in trapped ions \cite{Zhang2017}, and for such scenario the spin system holds long-range interactions across the entire network. 

At the initial time, we specify two regions in the network as shown in Fig.\ref{Fig1}.  As we have six spins in the system ($N=6$), we can assign three spins to each region as shown in Fig. \ref{Fig1} (b).  
The two regions are referred to as regions A (sites $0-2$) and region B (sites $3-5$), and we assume  these regions can be prepared in different initial states independently.
If both regions A and B are initially prepared in the same state, the total system would globally exhibit a quantum phase. For example, if the initial state globally breaks the discrete time-translational symmetry, the total system would be in the DTC phase.  Contrary to this, we prepare the region A in the DTC phase that breaks the discrete translational symmetry in time, while region B is prepared in another phase of matter.

\subsection{DTC growth and Effect of the dephasing}
We prepare the initial state  $ |111\rangle$ for region A to break the DTTS, where $|1\rangle_l$ is the eigenstate of the $\sigma_l^z$ at the $l$-th site. Next, a natural choice of the initial state for the region B is a state preserving the DTTS.  In this way, the system clearly exhibit two different phases in well-defined regions of the network.The aforementioned state for the region B is chosen to be $|{+++}\rangle$, hence the total initial state would be $|\psi(0)\rangle = |111\rangle_A \otimes |{+++}\rangle_B$, where $|+ \rangle_l = \left(|0\rangle_l + |1\rangle_l\right)/\sqrt{2}$. As the dynamics preserves the total number of excitations, the evolution of this initial state is restricted to subspaces with $3,4,5$ and $6$ excitations. To investigate the dynamics of the DTC, we calculate the local magnetizations $\langle \sigma_l^z(nT) \rangle$ at the stroboscopic times $t=nT$ for different dephasing rates $\gamma T =0, 0.012, 0.02$.
 
In Fig. \ref{Fig2} (a), we show the stroboscopic dynamics of the local magnetization $\langle \sigma_l^z (nT) \rangle $. The spin-spin interaction strength $J^{xy}_{lm}$ in the second Hamiltonian Eq.~\eqref{eq:Hamiltonian} allows the population transfer between regions A and B.   As a consequence, in a short time-scale, the $2T$-periodic magnetization or the DTC region initially located within the region A spreads to the region B.  This can be seen for all the dephasing rates. We refer to this spread of the DTC phase as DTC growth and the initial state for the region A as the DTC seed.  In the absence of the dephasing, the state remains pure at all times, and at the boundary of the spin network the population transfer reflects and causes interference.  In this situation, the magnetization dynamics does not reach a steady state. and we can observe the oscillation as shown in the left panel of Fig.~\ref{Fig2} (a).  On the contrary, when the dephasing is present, the state gradually loses the purity over time.  Thus, there is an interplay between the dissipation process and the coherent dynamics so as to  suppress interference.  As a result, the system is stabilized to the 2T-periodic dynamics spread over the entire system after a certain time.  Looking at the magnetization at each site, the DTC gradually grows over the entire network, starting from the DTC seed.  Generally, as the dephasing rate increases, the state losses its purity at a faster rate and reaches the steady state at shorter time scales. Thus, the number of periods (n) needed for the dynamics to stabilize decreases with the dephasing rate, as we can see by comparing the two left panels in Fig.~\ref{Fig2} (a).

Next, to understand the time evolution of the phase growth from the quantum state perspective, we calculate the negativity~\cite{Vidal2002} $N_B = \left(|| \hat{\rho}^{\Gamma_B}||_1-1\right)/2$ for each dephasing rate: $\gamma T =0, 0.012, 0.02$.  Here $\hat{\rho}^{\Gamma_B}$ is the partially transposed density matrix $\hat{\rho}$ with respect to subspace $B$ with $||\hat{X}||_1=\text{Tr}\sqrt{\hat{X}^{\dagger}\hat{X}}$ being the trace norm. Now, the negativity can be used as a measure of entanglement between the two regions A and B of the total system. The negativity for each dephasing rate is numerically calculated, and the results are shown in Fig.~\ref{Fig2} (b).  Initially, the negativity grows over time for all cases.  The entanglement between two regions increases via the interaction across the two regions.  When $\gamma T=0$, the negativity saturates after a certain period with slight fluctuations, though the magnetization dynamics would not be stabilized.  Contrary, if $\gamma T \neq 0$, the negativity peaks at a certain time and then begins to decrease to converge to zero due to the dephasing effect.   
By comparing both results in Fig.~\ref{Fig2} (a) and (b), we see that the DTC region grows via the interaction. and the DTC phase becomes stable as the system approaches to the steady state by dephasing.

\begin{figure}[t]
\centering
\includegraphics[width=0.48\textwidth]{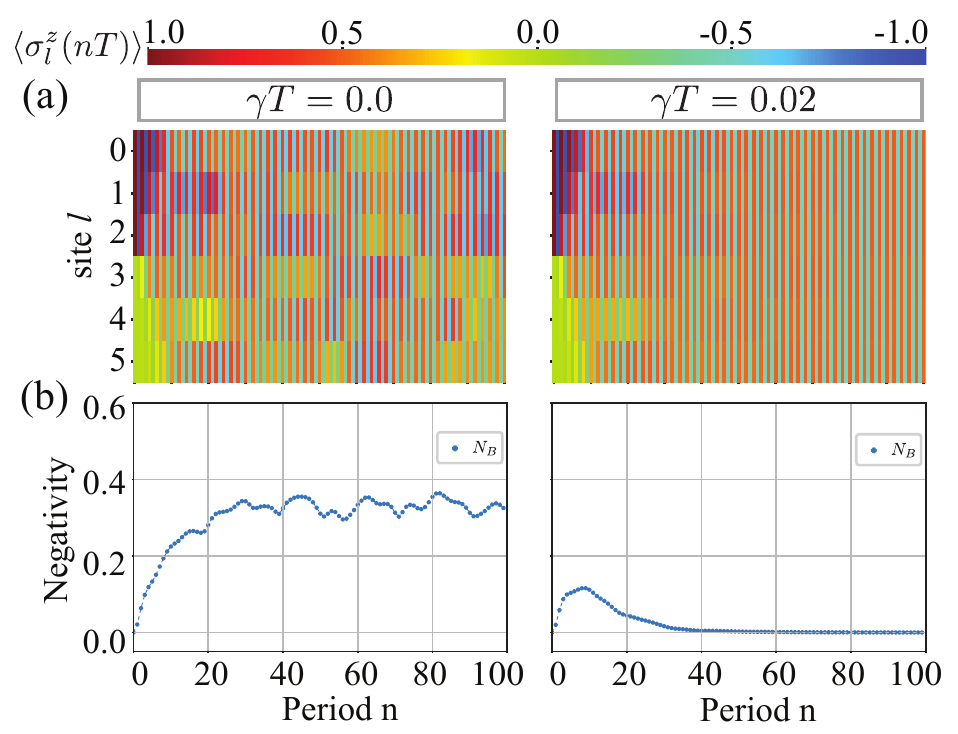}
\caption{ In (a) and (b), we show the results of the local magnetization $\langle \sigma_l^z(nT) \rangle$ and the negativity $N_B (nT)$ at periodic times $t=nT$ for a new initial state $|\phi \rangle = |111\rangle\otimes \hat{\rho}_B$, where $\hat{\rho}_B = I_{M_B}/{M_B}$ with the region B's Hilbert space size $M_B=2^3$ is a completely mixed state in region $B$. Here, we have chosen two dephasing rates $\gamma T =0, 0.02$, and have set up $J_0 T/2\pi=0.2$ with $\alpha=1.5$ and $WT/2\pi=0.0$.
}
\label{Fig3}
\end{figure}

In the above analysis, we have seen that the quantum transfer of excitations between the regions is essential for the quantum phase to be spread to the entire network as well as to be stabilized by the dephasing.  It is however not clear how these two competing factors contribute to generate a stable DTC. 
As the spread of the DTC seems to be related to the initial preparation of the system which we initially choose in the pure state $|\psi(0)\rangle = |111\rangle_A \otimes |+++\rangle_B$, what would happen region B starts with a mixed state. In that case the DTC region may not grow.  To investigate this, we consider the case where region B starts from the initially fully mixed state  ${\rho}_B (0) = I_{M_B}/{M_B}$ at the initial time, where $M_B = 2^{3}$ is the Hilbert space size of region B. 

In Fig.~\ref{Fig3}, we show the results of the dynamics of the local magnetization and the negativity for the new initial state with two different dephasing rates $\gamma T=0$, and $0.02$.  As shown in Fig.~\ref{Fig3} (a), despite the initial state in region B is the completely mixed state, we see that the DTC area grows over time in the first few periods. The entanglement between the two regions A and B increases in Fig~\ref{Fig3} (b). In the presence of dephasing, the DTC region spreads over the entire site and becomes stable, similar to the results in Fig~\ref{Fig2}. 
Additionally, to investigate the dependence of the initial state on DTC growth, we calculated the dynamics for different sizes of the DTC seed at the initial time, e.g. $|\psi(0)\rangle = |1\rangle \otimes |+++++\rangle$. It showed that the growth occurs in the magnetization dynamics as long as the seed exists in the initial state.

These results indicate that as long as we prepare a DTC seed in the initial state, the DTC region can grow via the interaction. Further, the dephasing in the dynamics plays an essential role on the DTC growth. We previously assumed that the rotational error is zero ($\epsilon= 0$), but we confirm the even with nonzero $\epsilon$, the numerical trend does no change, the dynamics shows DTC growth.

\subsection{The Liouvillian gap and DTC growth time scale}
As we have seen in the previous section, the DTC growth is induced by the interactions and the dephasing, and hence both the nature of the steady-state as well as the spectral properties of the Liouvillian operator $\hat{\hat{L}}^{\text{eff}}_{T} $ are intimately related to the DTC growth. 
To understand the DTC growth in more details, we investigate the structure of the Liouvillian operator. Since the effective Liouvillian operator $\hat{\hat{L}}^{\text{eff}}_{T} $ is time independent, using its eigenstates, the density matrix at stroboscopic times $t = nT$ can be written as
\begin{equation}
	|\hat{\rho}(nT) \rangle\rangle  = \left(\hat{\hat{\Phi}}_{T}\right)^{n} |\hat{\rho}(0) \rangle\rangle = \sum_l e^{\Lambda_l nT} c_l(0) | \Lambda_l^{R}\rangle\rangle.
	\label{eq:GenLiuDec}
\end{equation} 
Here, $c_l(0) = \langle\langle \Lambda_l^{L} |\hat{\rho}(0)\rangle\rangle$, and $| \Lambda_l^{L/R} \rangle\rangle$ are left/right eigenvectors of the effective Liouvillian operator $\hat{\hat{L}}^{\text{eff}}_{T} $ with the eigenvalue $\Lambda_l$~\cite{Minganti2018}. Here, we introduce an order among the eigenvalues as $\text{Re}(\Lambda_0) > \text{Re}(\Lambda_1)\ > \cdots > \text{Re}(\Lambda_{2^{2N}})$.  The steady state $|\hat{\rho}\rangle\rangle_{0}=|\hat{\rho}\rangle\rangle_{ss}$ satisfies
\begin{equation}
\hat{\hat{L}}^{\text{eff}}_{T}|\hat{\rho}\rangle\rangle_{ss} = 0.
\end{equation} 
with the eigenvalue $\Lambda_0 = 0$.  Thus, the relaxation time-scale $\tau \sim 1/\Delta$ (unit of $T$) can be estimated by calculating the Liouviullian gap~\cite{Kessler2012}
\begin{equation}
	\Delta = - \text{Re}(\Lambda_1),
\end{equation} 
which is the negative real part of the second largest eigenvalue $\Lambda_1$ of the Liouvillian operator.
Now, let us consider the DTC growth time scale ($\tau_{\text{DTC}}$) of our model. The time scale of the DTC growth $\tau_{\text{DTC}}$ should tell us the time for the DTC interference to be settled and stabilized, hence there are two steps in time crystal growth: the propagation of the DTC region and the DTC stabilization. The time scale of the first step can be determined by the propagation speed, which is related to the hopping strength as similar to the excitation transfer. The time scale of stabilization is related to the dissipation time of the system.   
The stabilization takes longer than the dynamical hopping rates.  Thus, we can estimate the time for the DTC growth from the stabilisation time scale, i.e. the Liouvillian gap is an appropriate measure to estimate $\tau_{\text{DTC}}$. 

\begin{figure}[t]
\centering
\includegraphics[width=0.48\textwidth]{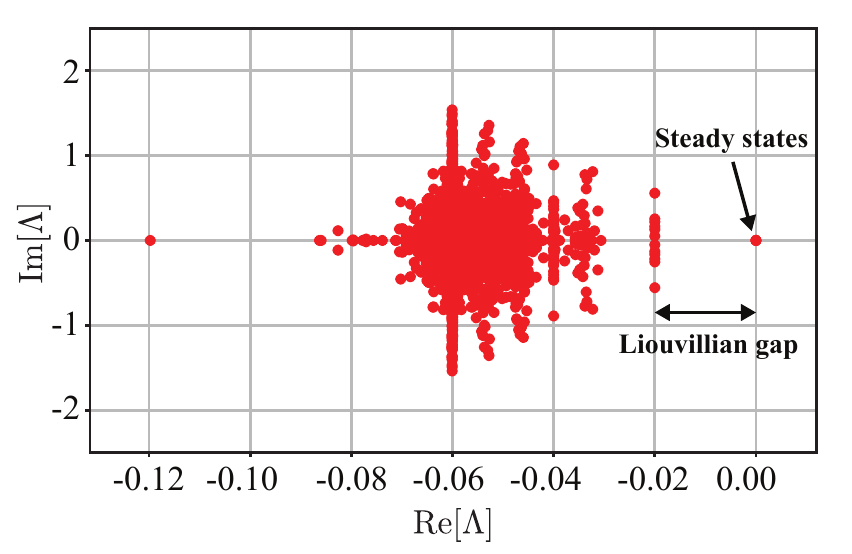}
\caption{
2T-effective Liouvillian operator $\hat{\hat{L}}^{2T}_{\text{eff}}$ structure for the spin network with $N=6$ : Red dots are complex eigenvalues $\Lambda$ of $\hat{\hat{L}}^{2T}_{\text{eff}}$. The system's parameters are the same as the rightmost case in Fig.~\ref{Fig2}. Because of the conservation of the total magnetization, there are multiple steady states.
}
\label{Fig4}
\end{figure}

Given the 2T-periodicity of the DTC, we use the effective Liouvillian operator $\hat{\hat{L}}^{2T}_{\text{eff}}$ . In general, this operator is given by
\begin{equation}
	 \hat{\hat{L}}^{\text{eff}}_{2T} =  \frac{1}{2T}\log{\left( \mathcal{T} e^{\int_0^{2T} \hat{\hat{L}}_
	 {\tau} d\tau} \right)} = \frac{1}{2T}\log{\left( \hat{\hat{\Phi}}_{T}^2\right)},
	\label{eq:2TeffLiouvillian}
\end{equation}
where we use the periodicity of the Liouvillian operator $\hat{\hat{L}}_t = \hat{\hat{L}}_{t+T}$ to perform the integration. When $\epsilon=0$, Eq.~\eqref{eq:2TeffLiouvillian} and Eq.~\eqref{eq:a-2TeffLiouvillian} give the same dynamical map. Then, the Liouvillian gap of the DTC growth is the second large eigenvalue of the Liouvillian operator $\hat{\hat{L}}_{2T}^{\text{eff}}$.

Now, let us look at the eigenvalues of the effective Liouvillian operator for the rightmost case in Fig.~\ref{Fig3} as an example. We show the numerical results in Fig.~\ref{Fig4}. Because our system preserves the total magnetization at even periods $t_n=2nT$, we find multiple steady-states~\cite{Vznidarivc2015}.  Since these steady states are not necessarily a thermal state, the DTC phase could exist when the system reaches the steady state. From this figure, we see that the Liouvillian gap is $\Delta = 0.02$ (unit of $1/T$), and the corresponding relaxation time is $\tau/T \sim 1/\Delta T = 50$. While we can estimate the DTC growth time scale to be around $\tau_{\text{CG}}/T = 40\sim 60$  period from Fig.~\ref{Fig3} (a) and (b), the estimated time-scale obtained by analizing the Liouvillian gap explains the time that the the DTC growth process take. 

\begin{figure}[t]
\centering
\includegraphics[width=0.48\textwidth]{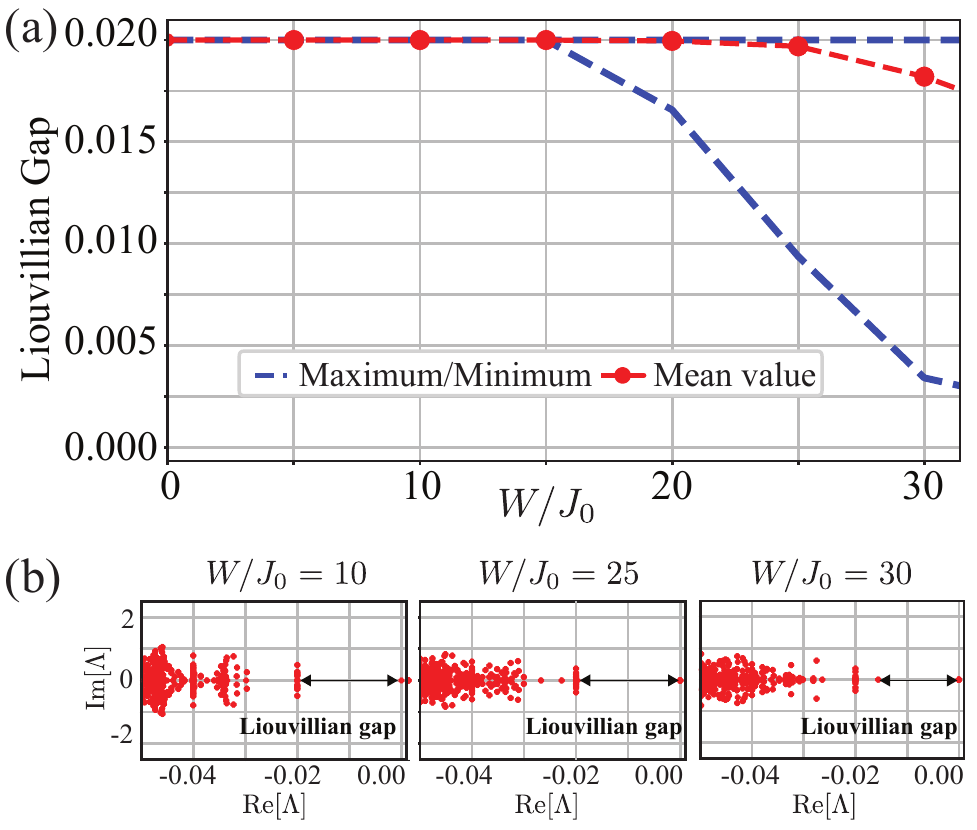}
\caption{ Liouvillian gap $\Delta$ of the spin chain with $N=6$ under the disorder. (a) Liouvillian gap for the several disorder strength $WT_2 = 0$ to the maximum value $\pi$. The red dots and dashed line are average value of the Liouvillian gap with 200 disorder realizations. The blue band is the width between maximum and minimum values of the Liouvillian gap. (b) Liouvillian structure of one disorder realization around the steady states for different strength of the disorder $W/J_0 = 10, 25$ and $30$. In both (a) and (b), we set $\gamma T = 0.02$, and $J_0 T/2\pi =0.2$ with $\alpha=1.51$.
}
\label{Fig5}
\end{figure}

\subsection{DTC growth vs Diagonal disorder}
We have previously seen that the DTC growth time scale in the absence of the diagonal disorder is well estimated by the Liouvillian gap. In the above case, the Liouvillian gap is equal to the dephasing rate $\gamma$, similarly to the case of a non-interacting spin chain under the dephasing effect~\cite{Shibata19}.   When there is strong diagonal disorder, the crystal growth might be suppressed due to Floquet Anderson localization (FAL)~\cite{Malishava2020}.  Here we investigate the time scale of the DTC growth in the presence of diagonal disorder by using Liouvillian gap.  In Fig.~\ref{Fig5} (a), we plot the average value of the Liouvillian gap over 200 disorder realizations for each value of the disorder strength.  In this figure, the blue lines represents the minimum and maximum values.

In the derivation of the Lindblad master equation\eqref{eq:Linovillian}, one assumes that the system is weakly coupled to a Markovian environment. In this derivation, we implicitly work in a regime where the dephasing rate is smaller than the characteristic energy scales of the system
~\cite{Gorini1976, Lindblad1976, Scopa2018, Manzano2020}. In our model, we work in a regime such that the Hamiltonian \eqref{eq:Hamiltonian} satisfies the aforementioned condition. However, in the presence of the strong disorder, the coupling effectively becomes smaller, and then the condition might be broken at stroboscopic times $t=2nT$(see Appendix \ref{appendixA}). Thus, there is a critical disorder strength $W^c_z$. For disorder strengths such that $W_z >W^c_z$ the effective coupling strength becomes smaller than the dephasing rate $\gamma$. Before and after the critical point, the Liouvillian spectrum and the Liouvillian gap undergo a drastic change~\cite{Shibata19}. The numerical results in Fig.~\ref{Fig5} (a) show that the DTC growth time-scale is not affected by the disorder up to a certain strength $W/J
_0 \sim 15$. Only in the strong disorder region, the effective coupling of the effective Hamiltonian $\hat{H}_{2T,W\neq 0}^{\text{eff}}$ becomes smaller than the dephasing rate, and the Liouvillian gap may also take a smaller value. When this happens, the DTC growth takes a longer time. 

Finally, let us take a closer look at the transition of the Liouvillian gap around the critical point.  We investigate the structure of the eigenvalues of the Liouvillian operator for the several disorder strengths.  We shows the results of the Liouvillian operator around the steady states for the different disorder strengths $W/J_0 = 10, 25$ and $30$ in Fig.\ref{Fig5} (b). We see that the structure around the gap does not change up to $W/J_0 \sim 15$.  However, when the disorder is stronger than this, the structure of the Liouvilian operator is gradually broken, and the gap gets smaller. As the dynamics tend to preserve the local magnetization in such a case, as a result, the growth of DTCs becomes slower.

\section{Conclusion}
In this paper, we have explored how a crystalline structure along the temporal axis grows on spin networks under the effect of dephasing. We illustrated the DTC growth for two different initial states: the pure state and the fully mixed state. By comparing the results from these two initial states, we showed that the purity of the initial state is not crucial, however a DTC seed is necessary for the DTC growth, and the dephasing is crucial for the system to form the DTC phase. 
We also analyzed the structure of the $2T$-effective Liouvillian operator and showed that the Lioucillian gap provides an indication of the DTC growth time scale. When diagonal disorder is present, the Liouvillian gap shows a critical point where its structure changes. 
Below this critical point, the gap is constant, despite the presence of the Anderson localization effect.
With the recent rapid progress in quantum control technology, our DTC growth experiment could be realized in several real quantum devices, including superconducting qubits and trapped ions. 

\section{Acknowledgments}
We thank T. Osada and J. Tangpanitanon for valuable discussions. This work was supported in part from the Japanese  MEXT  Quantum  Leap  Flagship  Program  (MEXT  Q-LEAP)  Grant  No.JPMXS0118069605, the MEXT KAKENHI Grant-in-Aid for Scientific Research on Innovative Areas Science of hybrid quantum systems Grant No.15H05870 and the JSPS KAKENHI Grant No. JP19H00662. 
 
\appendix
\section{Effective coupling with the strong disorder}
\label{appendixA}
 Here, we investigate how the coupling strength in the effective Hamiltonian changes in the presence of the disorder. To simplify the discussion, we consider a small system with two sites (N=2), as our example. In this model, the diagonal disorder effect is equal to the on-site energy gap between two sites. Thus, we consider the time-dependent Hamiltonian, as follows 
\begin{equation}
H^{(s)}(t) = 
\begin{cases}
\hbar g \left( \sigma_{1}^{x}  + \sigma_{2}^{x} \right) & 0 \leq t < T_1 \\
\hbar  J_0 \left( \sigma_{1}^{x} \sigma_{2}^{x} + \sigma_{1}^{y} \sigma_{2}^{y}  \right) + \hbar W \sigma_2^z & T_1 \leq t < T \ ,
\end{cases}
\end{equation}
where $2gT_1= \pi$. Because the total magnetization is preserved in our 2T-effective Hamiltonian, we only consider the difference of the on-site energy between two sites ($W^z_1, W^z_2$). Thus, the energy gap $W ( = W^z_2 - W^z_1)$ characterises the effect of the diagonal disorder. 

To obtain the $2T$-effective Hamiltonian, we begin with the square of the Floquet operator. Using the spin rotation properties, it can be written as
\begin{equation}
\hat{F}^2 = e^{ -i\left\{  J_0 \left( \sigma_{0}^{x} \sigma_{1}^{x} + \sigma_{0}^{y} \sigma_{1}^{y}  \right) +  W\sigma_2^z \right\} T_2} e^{-i\left\{  J_0 \left( \sigma_{0}^{x} \sigma_{1}^{x} + \sigma_{0}^{y} \sigma_{1}^{y}  \right) -  W \sigma_2^z \right\}T_2}.
\label{eq:2NFloquet}
\end{equation} 
As we discussed in the main text, $2T$-effective Hamiltonian preserves the total number of the excitations. Thus, to investigate the effect of the disorder on the effective couplings at stroboscopic times $t_n=2nT$, we consider a subspace for the one excitation. In this subspace, our system is a two level system, when we consider a the basis set
\begin{equation}
|10\rangle = \left( \begin{matrix}1\\0\end{matrix} \right),\quad
|01\rangle = \left( \begin{matrix}0\\1\end{matrix} \right),
\end{equation}
for the new subspace. In this basis,
the $2T$-effective Hamiltonian reads
\begin{equation}
	\hat{H}^{\text{eff}}_{2T} =  \left( \begin{matrix}
	\varepsilon_0 &  K\\
	 K^{*} & \varepsilon_1
	\end{matrix} \right),
\end{equation}
where $\varepsilon_{0/1}$ are real, and $K$ is a complex effective coupling, respectively. The effective hopping strength is characterized by the absolute value of the off-diagonal term $|K|$.  

\begin{figure}[t]
\centering
\includegraphics[width=0.46\textwidth]{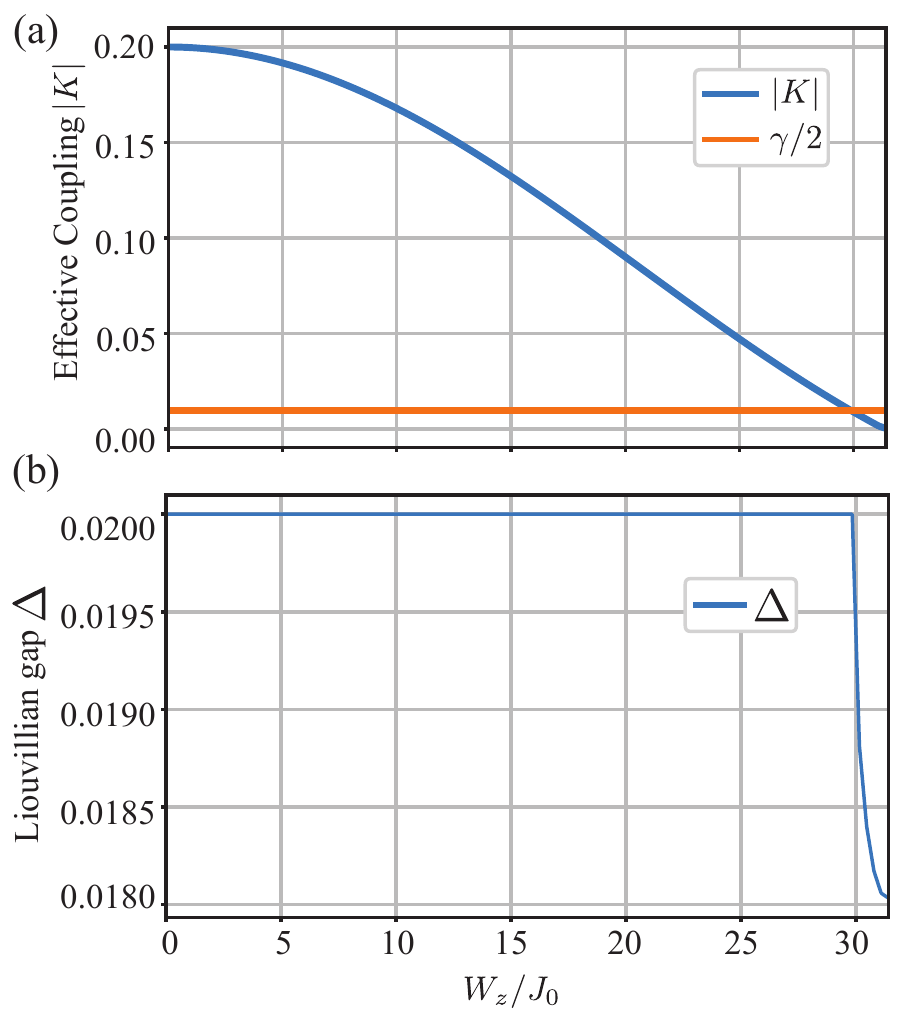}
\caption{ Effective coupling $|K|$ and Liouvillian gap $\Delta$ for a $N=2$ coupled spins under the effect of disorder. (a) The effective coupling $|K|$ for the several values of disorder strength from  $W_zT_2/2\pi= 0$ to $\pi$.  (b) Liouvillian gap for the several values of disorder strength from $ W_zT_2/2\pi = 0$ to $\pi$.  In both (a) and (b), we set $\gamma T = 0.02$, and $J_0 T/2\pi =0.2$.
}
\label{Fig1A}
\end{figure}

Now, within the one-excitation subspace, Eq.~\eqref{eq:2NFloquet} can be written as,
\begin{equation}
\hat{F}^2 = e^{-i\left(2J_0 \sigma^x + W\sigma^z \right)T_2} e^{-i\left(2J_0 \sigma^x -W \sigma^z  \right)T_2},
\end{equation}
where $\sigma^{\mu}$ ($\mu \in x, y, z$) are the Pauli matrices for the two level system.  First, in the absence of the disorder, the effective Hamiltonian is simply, 
\begin{equation}
	\hat{H}_{2T}^{\text{eff}} = \hbar J_0\sigma^x,
\end{equation}
Here, the hopping strength is $|K| = J_0$. Next, when the disorder strength is small ($WT < 1$), the effective Hamiltonian is approximately, 
\begin{equation}
	\hat{H}_{2T}^{\text{eff}} \approx \hbar J_0 \left(1 - \frac{ {W}^2 T^2}{6} \right) \sigma^x + \frac{J
	_0 WT}{2}\sigma^y,
\end{equation}
where we use the Baker-Campbell-Hausdorff formula at the lowest order, a method widely used to obtain the effective Hamiltonian~\cite{Estarellas19, Sakurai2021, Vajna2018}. Here, the strength is $|K| \sim J_0 \sqrt{1- W^2/12 }  < J_0 $, and its value is getting smaller than in the absence of the disorder. Finally, let us consider the case in which the disorder strength is sufficiently large $WT \gg 1$. In this case, we exactly derive the Hamiltonian, as follows,
\begin{equation}
	\hat{H}_{2T}^{\text{eff}} = - \frac{c \hbar \left( \sqrt{(2J_0)^2 +W^2} \sigma^x - W\sin(a) \sigma^y  \right) }{2T\sqrt{(2J_0)^2 \cos^2(a)+W^2}},
\end{equation}
where we use the group composition law of SU(2). Here the two parameters $a$ and $c$ are given by
\begin{equation}
\begin{split}
	a &= -T_2 \sqrt{(2J_0)^2 +W^2} \\
	\cos(c) &= 1 - \frac{2(2J_0)^2}{(2J_0)^2+W^2}\sin^2(a),
\end{split}
\end{equation}
respectively. Then, for the large gap, the effective Hamiltonian is approximately given by
\begin{equation}
	\hat{H}^{\text{eff}} _{2T}\approx \frac{2\hbar J_0}{WT}
	\left(
	\frac{\sin(2WT_2 )}{2}\sigma^x - \sin\left(WT_2\right)^2\sigma^y
	\right).
	\label{eq:EffHamiltonainWL}
\end{equation}
The corresponding coupling strength is $|K| \sim 2J_0 |\sin(WT_2)|/WT \ll J_0 $. Thus, as $W$ increases, the effective coupling strength decreases, and it gets smaller than the dephasing rage at the certain critical strength $W_c$. Form Eq.~\eqref{eq:EffHamiltonainWL}, the critical point is roughly estimated as $W_c \sim \pi/T_2(1-\gamma/(2J_0+\gamma))$. Thus, for the $N=2$ case, the critical point is in the strong disorder regime.

Now, to see the more details, we also numerically find the effective coupling $|K|$ for several gaps and show the result in Fig~\ref{Fig1A} (a). As the gap increases, the effective coupling $|K|$ decreases, and it crosses the value $\gamma$ around $W_c/J_0 \sim 29$, and the agreement between numerical and theoretical results are good. Here, we note that the critical point is bigger than that in the main text ($W_c/J_0 \sim 15$). We want to point out that in the model studied in the main text has more sites, long-range interactions and higher excitation-sectors that can make the critical point shift.

Finally, let us look a Liouvillian gap of the above model. Here, we consider the energy gap from $W T_2 /2\pi=0$  to $\pi$.  In Fig. \ref{Fig1A} (b), we show the  Liouvillian gap. We see that when the strength is smaller than the critical point of \ref{Fig1A} (a), the Liouvillian gap $\Delta$ is constant. Contrary, after the critical point, it becomes smaller due to the competition between dephasing and the effective coupling at stroboscopic times.

\end{document}